\documentclass[fleqn,usenatbib]{mnras}

\usepackage{newtxtext,newtxmath}

\usepackage[T1]{fontenc}
\usepackage{ae,aecompl}

\usepackage{graphicx}	
\usepackage{amsmath}	
\usepackage{amssymb}	

\title[The Acoustic Anomaly and the Lithium Problem]{Addressing the Acoustic Tachocline Anomaly and the Lithium Depletion Problem at the same time
}

\author[A. C. S. J\o rgensen \& A. Weiss]{
\and
Andreas Christ S{\o}lvsten J{\o}rgensen$^{1}$ \thanks{E-mail: acsj@mpa-garching.mpg.de}
and
Achim Weiss$^{1}$
\\
$^{1}$Max Planck Institut f\"ur Astrophysik, Karl-Schwarzschild-Strasse 1, 85748 Garching, Germany
}

\date{Accepted XXX. Received YYY; in original form ZZZ}

\pubyear{2018}

\begin{document}
\label{firstpage}
\pagerange{\pageref{firstpage}--\pageref{lastpage}}
\maketitle

\begin{abstract}
Current standard solar models neither account properly for the photospheric lithium abundance nor reproduce the inferred solar sound speed profile. Diffusive overshooting at the base of the solar convective envelope has previously been shown to solve either of these model inadequacies. In this paper, we present an analysis of solar models with four different parametrizations of diffusive overshooting. We find that these models are able to recover the correct lithium depletion, regardless of the parametrization, if overshooting is suppressed, during the early evolutionary stages. Further, parametrizations of diffusive overshooting have been shown to improve the inferred sound speed profile. However, none of the presented models are able to simultaneously solve both model inadequacies, showing that diffusive overshooting on its own is deficient to account for observations.
\end{abstract}




\section{Introduction}

Comparisons between observations and model predictions reveal several shortcomings of modern stellar structure models. Present model inadequacies include discrepancies between the predicted solar sound speed profile and the solar sound speed profile inferred from helioseismology \citep{jcd1985, jcd1991}, i.e. the study of solar oscillations \citep{Leighton1960, Ulrich1970}. Especially near the tachocline, the sound speed difference between the Sun and current standard solar models show a striking anomaly. Lithium depletion in stellar convective envelopes is another issue that still has to be properly accounted for: meteoritic measurements and spectroscopic measurements of the solar photosphere show that the present solar abundance of $^{7}\mathrm{Li}$ is roughly 100-200 times lower than the initial one \citep[][]{Greenstein1951, Asplund2009}. Standard solar models do not reproduce this depletion. Moreover, surveys reveal that the Sun is not exceptional for a star of its age, mass and metallicity: solar-like stars are generally lithium poor \citep[e.g.][]{Baumann2010}.

The tachocline anomaly was immediately apparent from the first inversions of helioseismic data \citep[][]{jcd1985,jcd1988}. This discrepancy in the sound speed profile may partly reflect errors in the opacities \citep{jcd2010} but can also be addressed, by including additional mixing beyond the convective boundaries of stellar models \citep{Elliott1998,jcd1993,Richard1996}. This additional mixing affects the composition and hereby alters the sound speed, potentially getting rid of the tachocline anomaly. Other seismic properties, such as the mode amplitudes, can be employed to impose further restrictions on the mixing processes \citep{jcd2011}.

Mixing below the solar convective envelope furthermore transports lithium to hotter layers, where it is depleted in thermonuclear reactions, solving the lithium depletion problem. This solution was originally suggested by \cite{Boehm1963} and does indeed deplete lithium sufficiently to match the present solar photospheric values, if the parameters involved are adequately tuned. That being said, while observations \citep[e.g.][]{Chaboyer1998} indicate that this depletion takes place on the main sequence (MS), solar models with overshooting deplete lithium prematurely \citep[e.g.][]{Ahrens1992}, unless the suggested additional mixing is somehow inhibited on the pre-main sequence \citep[pre-MS,][]{Schlattl1999}. We will elaborate upon this in Section~\ref{sec:lith}.

Consequently, as discussed by several authors, both of the issues raised above can be {\color{black}addressed} simultaneously, by including mixing in the radiative zone \citep[][]{jcd1993,Richard1996, Schlattl1999, Andrassy2013}.

Several processes have been advocated to contribute to the necessary additional mixing, including differential rotation between the radiative and the convective zones \citep[e.g.][]{Spiegel1992}, internal gravity waves \citep[e.g.][]{Garcia1991}, and the penetration of convective plumes into the radiative zone, so-called overshooting. Different authors offer different pictures and models for the latter; for instance, based on \cite{Spruit1997}, \cite{Andrassy2013,Andrassy2015} argue for overshooting, assuming that convective settling beyond the convective boundary can be understood in terms of an entropy difference between the sinking material and its surroundings. 

One way of taking overshooting into account amounts to extending the adiabatic region by a fraction of a pressure scale height into the subadiabatic region beyond the base of the convection envelope. Such quasi-adiabatic penetration can be achieved, using mixing-length theory \citep{Zahn1991}. However, just as it is the case for models without overshooting, this approach gives rise to an abrupt transition in the temperature gradient, which is neither favoured by helioseismic measurements nor by hydrodynamical simulation of the convective plumes. Other treatments of overshooting, on the other hand, do not suffer from this deficit \citep[e.g.][]{Andrassy2013}. For a detailed discussion of this topic, we refer to \cite{jcd2011}.

Yet another method to take the structural effects of overshooting into account is to adjust the temperature gradient in the overshooting region, drawing from helioseismology and the overshooting profiles of 3D simulations \citep{jcd2011}. Finally, one may introduce additional mixing, by altering the diffusion coefficients in the overshooting layer \citep[e.g.][]{Baraffe2017,jcd2018,Schlattl1999}. Such adjustments of the diffusions coefficient may be justified by the attempt to mimic the outcome of 2D or 3D hydrodynamic simulations \citep{Freytag1996,Pratt2016}.

In this paper, we follow the last approach mentioned above, and hence restrict ourselves to diffusive overshooting. While the diffusion coefficients are altered to accommodate for additional mixing in the radiative zone, the radiative temperature gradient is used within the overshooting layer.

We investigate four different parametrizations of the diffusion coefficients in the overshooting layer, restricting ourselves to the solar case. The aim is to evaluate, whether each of these parametrizations can solve the lithium  depletion problem, and whether models that succeed simultaniously solve the tachocline anomaly. To this end, we have performed solar calibrations for each overshooting parametrization, using the Garching Stellar Evolution Code, \textsc{garstec} \citep{Weiss2008}. All solar models have a radius $R_\odot$ of $6.95508\times 10^{10}\,\mathrm{cm}$, a total luminosity $L_\odot$ of $3.846\times10^{33}\,\mathrm{erg \,s^{-1}}$, and an age of $t_\odot$ of $4.57\,\mathrm{Gyr}$. We employ the OPAL equation of state \citep[EOS, ][]{OPAL2005a,OPAL2005}. At low temperatures we extend the EOS with the EOS by \cite{Hummer1988}. We use either the composition suggested by \cite{Grevesse1998} (GS98) or by \cite{Asplund2009} (AGSS09) with the corresponding OPAL opacities \citep{Ferguson2005,Iglesias1996} and surface abundance $Z_\mathrm{s}$ of heavy elements relative to the surface abundance $X_\mathrm{s}$ of hydrogen. In all cases, we use the reaction rates suggested by \cite{Adelberger2010}. {\color{black}We include microscopic diffusion of H, $^{3}$He,  $^{4}$He, $^{12}$C, $^{13}$C, $^{14}$N, $^{15}$N, $^{16}$O, $^{17}$O, $^{20}$Ne, $^{24}$Mg, and $^{28}$Si that are considered in the nuclear reaction network \citep{Weiss2008}. $^7$Li and $^9$Be are likewise diffused but are treated as tracer elements.} Moreover, we have only implemented overshooting at the bottom of the convective envelope. We note that overshooting at the core must generally be modelled differently from envelope overshooting \citep[e.g.][]{Schlattl1999}. {\color{black}In the case of solar calibration models, core overshooting is only relevant on the pre-MS.}

\section{Overshooting Approaches} \label{sec:2}

A very simplistic diffusive approach is to extend the convective zone by an overshooting layer, for which the diffusion coefficient $D_\mathrm{ov}$ is assumed to be a step function:
\begin{equation}
D_\mathrm{ov}(r) = D_0, \quad \mathrm{for \; \;} r_\mathrm{cz} - l_\mathrm{ov} < r < r_\mathrm{cz}. \label{eq:constD}
\end{equation}
In the following, we use the label CON to refer to models, for which the diffusion coefficient is constant within the overshooting layer, i.e. for which the parametrization given by Eq.~(\ref{eq:constD}) is employed. The default case of no overshooting is labelled DEF.

In Eq.~(\ref{eq:constD}), $r$ denotes the distance from the stellar centre, and $r_\mathrm{cz}$ denotes the radius of the lower convective boundary, i.e. the radius below which the Schwarzschild criterion of convective instability, $\nabla_\mathrm{rad}>\nabla_\mathrm{ad}$, no longer holds true. $D_0$ takes the value of the diffusion coefficient half a scale height above the convective boundary, as the diffusive velocity is formally zero at the boundary. Hence, based on mixing length theory, $D_0$ is $\frac{1}{3}\ell_\mathrm{mix}v_\mathrm{MLT}$. Here, $\ell_\mathrm{mix}$ and $v_\mathrm{MLT}$ are the mixing length and the convective velocity, respectively. The units are hence $\mathrm{cm^2 \, s^{-1}}$. The width $l_\mathrm{ov}$ of this overshooting layer may be chosen freely to fit observations.

\cite{Freytag1996} and \cite{Bloecker1998} have suggested a more sophisticated approach, based on two-dimensional hydrodynamical simulations. Here, we employ the implementation by \cite{Schlattl1999}, according to which the diffusion coefficient in the overshooting region~is
\begin{equation}
D_\mathrm{ov}(r) = D_0\exp\left( \frac{-2(r_\mathrm{cz}-r)}{f_\mathrm{ov} H_p} \right). \label{eq:bloecker}
\end{equation}
In the following, we use the label FOV to refer to the overshooting parametrization by \cite{Freytag1996} given by Eq.~(\ref{eq:bloecker}).

In Eq.~(\ref{eq:bloecker}), $f_\mathrm{ov}$ is a free parameter, and $D_0$ is defined as above. Finally, $H_p$ denotes the pressure scale height at the base of the convection zone, including a geometrical restriction: If $2H_p$ exceed the extent of the convection zone ($\Delta r_\mathrm{cz}$), $H_p$ is adjusted by a factor of $(\Delta r_\mathrm{cz}/(2H_p))^2$. This geometrical limitation of overshooting is especially important, when considering core overshooting but is mostly irrelevant in the present case. We refer to \cite{Higl2017}, who have modelled detached eclipsing binaries, for a discussion of a case for which the implementation of this geometrical cut-off becomes essential. We apply this geometrical cut-off for all diffusive overshooting parametrizations presented in this paper.

The seismic implications of the overshooting parametrization by \cite{Freytag1996} and the associated lithium depletion have previously been discussed by \cite{Schlattl1999}. While they also address cases, where core overshooting is taken into account, we only include envelope overshooting, in order to facilitate a meaningful comparison between the different approaches. We introduce a cutoff, when $D_\mathrm{ov}(r)$ becomes smaller than $10^{-20}D_0$.

Generally speaking, exponentially decaying diffusive mixing is often used to improve stellar models \citep[e.g.][where the scaleheight is determined by the density]{Miglio2007, Buldgen2017}.

Recently, \cite{Pratt2016} and \cite{Baraffe2017} have suggested yet another overshooting parametrization, based on two-dimensional hydrodynamical simulations of compressible convection in young stellar objects (YSOs):
\begin{equation}
D_\mathrm{ov}(r) = D_0\left\lbrace  1 - \exp \left[ -\exp \left( \frac{-\frac{(r_\mathrm{cz}-r)}{R}-\mu}{\lambda} \right) \right]   \right\rbrace. \label{eq:pratt}
\end{equation}
We attribute the label PB to this parametrization by \cite{Pratt2016} and \cite{Baraffe2017}. In accordance with \cite{Baraffe2017}, we adopt $\mu = 5\times10^{-3}$ and $\lambda = 6\times10^{-3}$, based on the simulation of a $1\,M_\odot$ pre-MS star by \cite{Pratt2016}. Assuming that processes, such as rotation, may restrict the overshooting of convective plumes, \cite{Baraffe2017} introduce a limiting width $d_\mathrm{ov}$ of the overshooting region\footnote{{\color{black}In order to avoid extensive lithium depletion on the pre-MS, \cite{Baraffe2017} set $d_\mathrm{ov}$ to $0.1\,H_\mathrm{p}$, when the rotation rate exceeds a ceterain critical value. Below this critical rotation rate, $d_\mathrm{ov}$ is set to $1\,H_\mathrm{p}$. For further details we refer to Section~\ref{sec:lith} and the quoted paper.}}. While \cite{Baraffe2017} discuss the issue of depletion, they do not address the asteroseismic implications of Eq.~(\ref{eq:pratt}).

Finally, \cite{jcd2007} and \cite{jcd2018} adopt an additional diffusive process that follows a power-law:
\begin{equation}
D_\mathrm{ov}(r) = D_\textsc{jcd} \left( \frac{v-v_0}{v_\mathrm{c}-v_0} \right)^\alpha, \quad v=\frac{1}{\rho}. \label{eq:jcdD}
\end{equation}
We refer to this last parametrization as JCD. Here $\alpha$, $v_0$ and $D_\textsc{jcd}$ are free parameters, while $v_\mathrm{c}=1/\rho_\mathrm{c}$, where $\rho_\mathrm{c}$ is the density at the base of the convection zone. In order to obtain sensible seismic results, using the solar composition recommend by AGSS09, \cite{jcd2018} have implemented the opacity correction suggested by \cite{jcd2010}. We have not included this correction. While \cite{jcd2018} address the helioseismic implications of their diffusive overshooting parametrization, they do not investigate the resulting lithium depletion. 

In accordance with \cite{jcd2018}, we adopt $\alpha = 4.25$ and $v_0 = 0.15\,\mathrm{g^{-1}cm^3}$ and introduce a cutoff at $v=v_0$. If $D_\mathrm{ov}(r)$ becomes smaller than $10^{-20}D_\mathrm{ov}(r_\mathrm{cz})$, we likewise introduce a cutoff. We have varied $D_\mathrm{\mathrm{JCD}}$, including the value employed by \cite{jcd2018}: $D_\mathrm{\mathrm{JCD}}=150\,\mathrm{cm^2 \, s^{-1}}$.

\section{Lithium Depletion} \label{sec:lith}

We have adjusted the free parameters $l_\mathrm{ov}$, $f_\mathrm{ov}$, $d_\mathrm{ov}$, and $D_\mathrm{\mathrm{JCD}}$, in order to obtain a lithium depletion by a factor of 100-200, using the four different overshooting parametrizations described in Section~\ref{sec:2}.

The associated overshooting parameters can be found in the text below together with a discussion of each case. The relative depletion of lithium until the present solar age according to each model is listed in Table~\ref{tab:liseis} alongside other model properties. These include $r_\mathrm{cz}$ as well as the corresponding acoustic depth:
\begin{equation}
\tau_\mathrm{cz} = \int_{r_\mathrm{cz}}^{R_{\odot}} \frac{\mathrm{d}r'}{c},
\end{equation}
where $c$ denotes the sound speed. Moreover, the table includes the mixing length parameter\footnote{The value of the mixing length parameter depends on the boundary conditions used in the solar model calibration \citep[e.g.][]{Weiss2008}. We have here consistently used Eddington grey atmospheres.} \citep[$\alpha_\textsc{mlt}$][]{Boehm1958}, to illustrate the influence of the overshooting scheme on the stellar parameters. Finally, the helium mass fraction $Y_\mathrm{s}$ at the solar surface is listed. For all models in Table~\ref{tab:liseis} that lead to a satisfactory depletion of lithium, $Y_\mathrm{s}$ is roughly within $2\sigma$ of the value recommended \cite{BasuAntia2004}: $0.2485\pm0.0035$. The model without overshooting is in slightly worse agreement with the observed value of $Y_\mathrm{s}$. However, this conclusion depends on the input physics, and very good agreement without overshooting can be obtained \citep[cf.][]{Weiss2008}. {\color{black}The variation in $Y_\mathrm{S}$ that results from the inclusion of overshooting reflects the requirements set by the calibration procedure: the inclusion of overshooting effectively extends the convective envelope and thereby influences the chemical profile --- and thus the surface composition. At the same time, the solar calibration models are all required to reproduce the solar radius ($R_\odot$), the solar luminosity ($L_\odot$), and the solar surface abundance of heavy elements ($Z_\mathrm{S}/X_\mathrm{S}$). This is achieved by altering the the mixing length (see Tab.~\ref{tab:liseis}, last column) and the initial composition, which likewise affects $Y_\mathrm{S}$. Thus, the depth of the convective envelope and the surface helium abundance change, due to the regulating effect of the calibration procedure. The various overshooting schemes lead to 
slight variations only as a second order effect.}

\begin{table}
 \caption{Properties of the models presented in Fig.~\ref{fig:lith}, using GS98. $\mathrm{Li_0/Li_\odot}$ denotes the initial model surface lithium abundance ($N(\mathrm{Li})/N(\mathrm{H})$) relative to the predicted present one. $\Delta r_\mathrm{ov}$ refers to the width of the overshooting layer. As regards the JCD parametrization described by Eq.~(\ref{eq:jcdD}), $D_\textsc{jcd}=150 \, \mathrm{cm^2 \, s^{-1}}$ in case~a, while $D_\textsc{jcd}$ is two magnitudes higher in case~b. The values for $D_\textsc{jcd}$ and the remaining free parameters can be found in the text. $\alpha_\textsc{mlt}$ denotes the mixing length parameter. The initial abundance of helium and heavy elements predicted by the calibration lie in the intervals $0.265-0.269$ and $0.0178-0.0187$, respectively. All models in this table take microscopic diffusion of metals into account.
 }
 \label{tab:liseis}
 \begin{tabular}{lllllll}
  \hline
  $D_\mathrm{ov}$ & $\mathrm{Li_0/Li_\odot}$ & $\frac{\Delta r_\mathrm{ov}}{R_\mathrm{\odot}}$ & $\frac{r_\mathrm{cz}}{R_\mathrm{\odot}}$ & $\mathrm{\tau_\mathrm{cz}}$ [s] & $Y_\mathrm{S}$ & $\alpha_\textsc{mlt}$ \\
  \hline
  DEF  & 3.7 &  --- & 0.715 & 2083 & 0.2393 & 1.80 \\[2pt]
  CON  & $1.2\times10^2$ & 0.08 & 0.718 & 2074 & 0.2451 & 1.77 \\[2pt]
  FOV & $1.7\times10^2$ & 0.16 & 0.718 & 2073 & 0.2454 & 1.77 \\[2pt]
  PB  & $1.6\times10^2$ & 0.08 & 0.718 & 2074 & 0.2450 & 1.77 \\[2pt]
  JCD, a & 4.0 & 0.36 & 0.717 & 2078 & 0.2415 & 1.79 \\[2pt]
  JCD, b & $1.4\times10^2$ & 0.36 & 0.719 & 2072 & 0.2465 & 1.77 \\[2pt]
  \hline
 \end{tabular}
\end{table}

When computing $D_\mathrm{ov}(r)$, using Eqs~(\ref{eq:constD})-(\ref{eq:pratt}), lithium is sufficiently depleted. {\color{black}However, this depletion takes place on the pre-MS, as already discussed by \cite{Baraffe2017} and \cite{Schlattl1999}}. This is inconsistent with observations (of e.g. the Pleiades or the Hyades), according to which the depletion takes place on the MS at a rate that depends on the angular momentum \citep[cf. e.g.][]{Chaboyer1998,Jones1997,Jones1999}.

\cite{Ventura1998} propose that the inhibition of the lithium depletion on the pre-MS may be explained by rotationally induced magnetic fields --- the authors also address the influence of changes in other input physics, such as the composition. Based on numerical studies \citep[cf.][]{Ziegler2003,Brummell2007,Brun2017}, \cite{Baraffe2017} equally argue that rotation limits the convective plumes, reducing $d_\mathrm{ov}$ at high angular velocities. Hence \cite{Baraffe2017} evaluate the rotation period, using Kawaler's law \citep{Kawaler1988,Bouvier1997,Viallet2012}, and assume $d_\mathrm{ov}$ to be a step function of the angular velocity. This necessitates the introduction of {\color{black}additional} parameters that are neither restricted by simulations nor by observations. That being said, for many choices of these parameters, their method effectively amounts to introducing one additional free parameter: the time $t_\mathrm{ov}$, before which overshooting is negligible or, at least, relatively inefficient.

Setting $t_\mathrm{ov}$ to {\color{black}the first} $10^8\,\mathrm{yrs}$ and changing $d_\mathrm{ov}$ from $0.10H_p$ to $0.99H_p$ at $t_\mathrm{ov}$, we obtain models that deplete Lithium {\color{black}predominantly} on the MS, when using GS98 and the PB parametrization. These parameter values are overall consistent with the results presented by \cite{Baraffe2017}, who effectively use a higher value of $t_\mathrm{ov}$.

Applying the same arguments regarding the suppression of overshooting during the early evolutionary stages, we have also introduced the parameter $t_\mathrm{ov}$, when employing the CON or FOV parametrization, i.e. Eq. (\ref{eq:constD}) or (\ref{eq:bloecker}). In accordance with \cite{Schlattl1999}, who likewise initialized overshooting on the ZAMS, this allows us to compute models, for which the main lithium depletion takes place on the MS. 

When employing the CON or FOV parametrization, we set $t_\mathrm{ov}=10^8\,$yr, $l_\mathrm{ov}=1.03$, and $f_\mathrm{ov}=0.083$, respectively, and neglect overshooting completely before $t_\mathrm{ov}$. These parameter values seem qualitatively consistent with \cite{Schlattl1999}, according to whom a satisfactory lithium depletion can be obtained, using $f_\mathrm{ov}=0.07$, when starting envelope overshooting on the ZAMS and ignoring core overshooting\footnote{We note that \cite{Schlattl1999} use $Z_\mathrm{s}/X_\mathrm{s}=0.0245$, as recommended by \cite{Grevesse1993}. Adopting the same surface metallicity, we find that lower values of $l_\mathrm{ov}$, $f_\mathrm{ov}$, $d_\mathrm{ov}$, and $D_\textsc{jcd}$ are needed to reach the same lithium depletion than in the case of $Z_\mathrm{s}/X_\mathrm{s}=0.0230$.}.

\begin{figure}
\centering
\includegraphics[width=\linewidth]{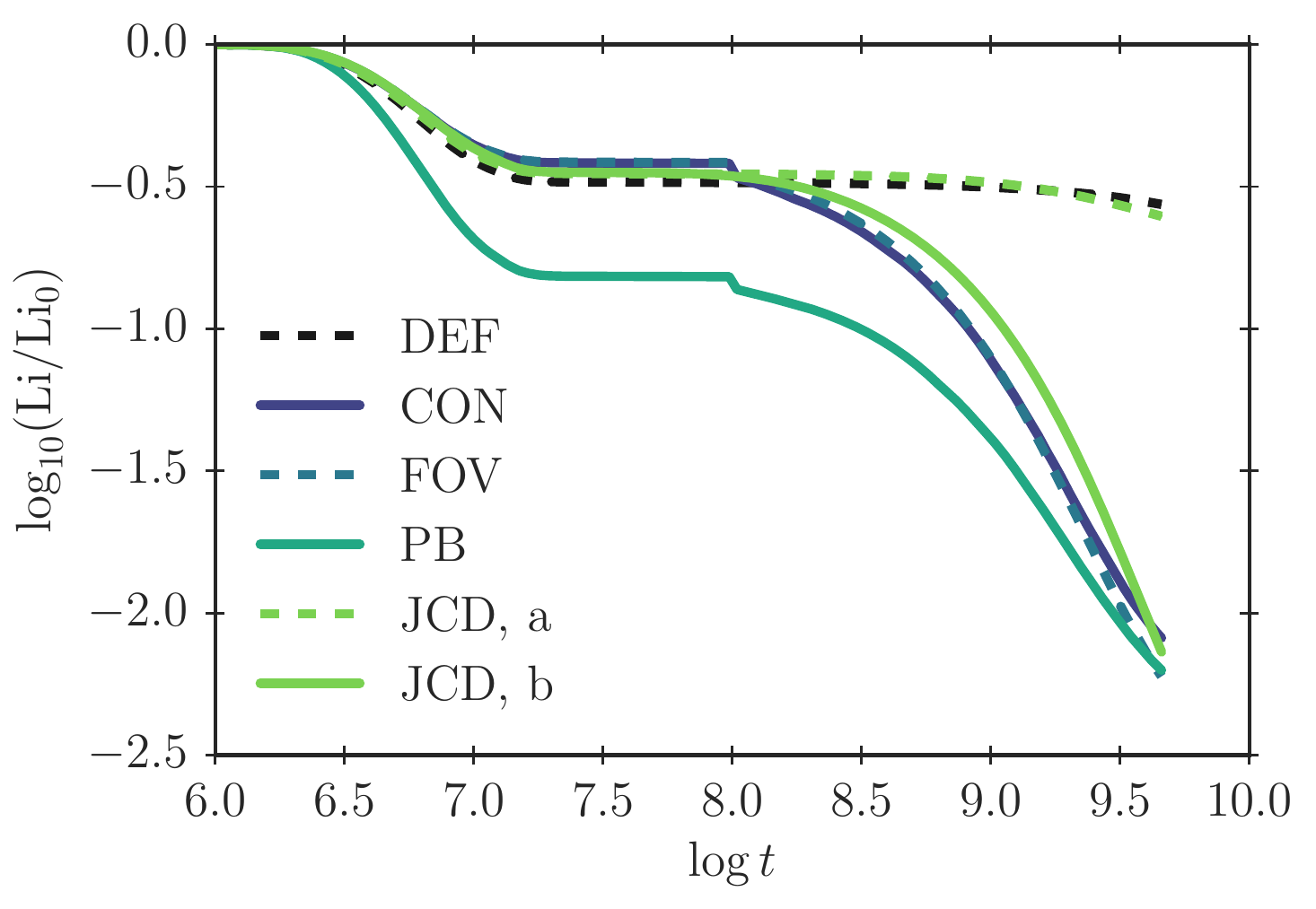}
\caption{Relative change in $N(\mathrm{Li})/N(\mathrm{H})$ at the solar surface surface as a function of time ($t$) for solar models, using GS98 and four different overshooting parametrizations, as well as a standard solar model, for which overshooting is neglected. $\mathrm{Li_0}$ denotes the initial surface lithium abundance: $A(\mathrm{Li_0})=3.3$.
}
\label{fig:lith}
\end{figure}

In the case of the JCD parametrization, i.e. Eq.~(\ref{eq:jcdD}), overshooting only depletes the surface lithium abundance by factor of four (cf.~case~a in Table~\ref{tab:liseis}), when using the parameter values suggested by \cite{jcd2018}. We therefore vary $D_\textsc{jcd}$. To obtain a lithium depletion by a factor of 100-200, we need to increase $D_\textsc{jcd}$ by two orders of magnitude. Fig.~\ref{fig:lith} shows the results for\footnote{For this choice of $D_\textsc{jcd}$, $\rho_\mathrm{c}$ in Eq.~(\ref{eq:jcdD}) is $0.1756\,\mathrm{g\,cm^{-3}}$ at the age of the present Sun. When using $D_\textsc{jcd} = 150 \, \mathrm{cm^2 \, s^{-1}}$, $\rho_\mathrm{c}=0.1789\,\mathrm{g\,cm^{-3}}$ for the solar calibration model. For comparison, $\rho_\mathrm{c}=0.1902\,\mathrm{g\,cm^{-3}}$ for the modified Model~S \citep{jcd1996}, used by \cite{jcd2018}. We have rerun the calculations keeping $\rho_\mathrm{c}$ fixed, using the value from \cite{jcd2018}. While this affects the choice of $D_\textsc{jcd}$, we reach the same qualitative conclusions as drawn in the present paper.} $D_\textsc{jcd} = 1.1\times 10^4 \, \mathrm{cm^2 \, s^{-1}}$ (cf. case~b in Table~\ref{tab:liseis}). Although $D_\textsc{jcd}$ is several orders of magnitudes lower than $D_0$, a high lithium depletion is achieved, as the overshooting layer is significantly deeper than for the other three parametrization schemes. 

In contrast to the other overshooting parametrizations presented in this paper, no additional parameter is needed to prevent a significant lithium depletion during the early evolutionary stages, in the case of the JCD parametrization. Here, we have set\footnote{\cite{jcd2018} start their simulation on the ZAMS. For the sake of consistency, one may hence choose {\color{black}a finite} $t_\mathrm{ov}$. We have therefore repeated the computations for the JCD parametrization, setting $t_\mathrm{ov} = 10^7\,$yr and $10^8\,$yr. The corresponding change in the depletion factor of the surface lithium abundance is $\lesssim 10\,$\%.} $t_\mathrm{ov} = 0$. 

For the CON, FOV, and PB parametrization, the relevant time scale for diffusion,
\begin{equation}
\tau_\mathrm{ov} \sim \frac{\Delta r_\mathrm{ov}^2}{\langle D_\mathrm{ov}(r) \rangle} = \frac{\Delta r_\mathrm{ov}^3}{\int D_\mathrm{ov}(r) \mathrm{d}r},
\end{equation}
in the overshooting region is of the order of {\color{black}decades} or shorter, throughout the solar evolution up until the present solar age. In the case of the JCD parametrization, for $D_\textsc{jcd}=150 \, \mathrm{cm^2 \, s^{-1}}$, $\tau_\mathrm{ov}>7.8\,$Gyr during the entire evolution, quickly exceeding the age of the universe by $1-2$ orders of magnitude. This clarifies, why the parameter values suggested by \cite{jcd2018} do not lead to a sizeable lithium depletion. When increasing $D_\textsc{jcd}$ by two orders of magnitude, $\tau_\mathrm{ov}>1.3\,$Myr during the entire evolution, reaching billions of years and exceeding the age of the universe during most of the evolution. This rationalizes, why the parameter $t_\mathrm{ov}$ is obsolete for this prescription. {\color{black}The longer time scale associated with the JCD parametrization reflects the low value of $D_\textsc{JCD}$ relative to $D_0$, the decrease of the diffusion coefficient with depth and the high penetration depth}. {\color{black}As already noted above,} the longer time scale is balanced by the higher temperatures reached by the overshooting material, due to the deeper penetration depth.

Figure~\ref{fig:lith} summarizes the evolution of the surface lithium abundance for all models described above. Here, we compare the initial number density $N(\mathrm{Li)}$ of Lithium relative to the number density $N(\mathrm{H})$ of hydrogen with the final one, since the abundance $A(\mathrm{Li})$, to which the literature refers, is $\log_{10}\left[N(\mathrm{Li})/N(\mathrm{H})\right]+12$.

It is worth noting that we {\color{black}found} the lithium depletion to be rather sensitive to the input physics, the overshooting parameters and the other model parameters, {\color{black}by computing several additional solar calibration models}. This includes the opacities, the element diffusion and the bulk composition. We attribute this sensitivity to the extreme temperature dependence of the lithium burning rate.

\begin{table}
\caption{Same as Table~\ref{tab:liseis} but for AGSS09. $A(\mathrm{Li_0})=3.2$. The initial abundances of helium and heavy elements predicted by the calibration lie in the intervals $0.259-0.265$ and $0.0140-0.0148$, respectively. All models in this table take microscopic diffusion of metals into account.
}
 \label{tab:liseis2}
 \begin{tabular}{lllllll}
  \hline
  $D_\mathrm{ov}$ & $\mathrm{Li_0/Li_\odot}$ & $\frac{\Delta r_\mathrm{ov}}{R_\mathrm{\odot}}$ & $\frac{r_\mathrm{cz}}{R_\mathrm{\odot}}$ & $\mathrm{\tau_\mathrm{cz}}$ [s] & $Y_\mathrm{S}$ & $\alpha_\textsc{mlt}$ \\
  \hline
  DEF & 2.5 & --- & 0.725 & 2051 & 0.2338 & 1.79 \\[2pt]
  CON & $1.3\times10^2$ & 0.10 & 0.728 & 2040 & 0.2407 & 1.76 \\[2pt]
  FOV & $1.5\times10^2$ & 0.19 & 0.728 & 2040 & 0.2410 & 1.75 \\[2pt]
  PB & $1.3\times10^2$ & 0.10 & 0.728 & 2040 & 0.2406 & 1.76 \\[2pt]
  JCD, a & 2.6 & 0.37 & 0.726 & 2048 & 0.2361 & 1.78 \\[2pt]
  JCD, b & $1.5\times10^2$ & 0.37 & 0.729 & 2038 & 0.2421 & 1.75 \\[2pt]
  \hline
 \end{tabular}
\end{table}

To illustrate this sensitivity to the input physics, we have repeated the calculations for AGSS09. We set $Z_\mathrm{s}/X_\mathrm{s}=0.0179$. The results are summarized in Table~\ref{tab:liseis2}. Here, we likewise use $t_\mathrm{ov} = 10^8\,$yr but adjust the remaining overshooting parameters, in order to obtain the required lithium depletion: we set $l_\mathrm{ov}=1.26$, $f_\mathrm{ov}=0.103$ and $D_\textsc{jcd}=2.5\times 10^4\,\mathrm{cm^2 \, s^{-1}}$. In the case of the PB parametrization, $d_\mathrm{ov}$ switches from $0.1H_p$ to $1.23H_p$ at $t_\mathrm{ov}$.

Again, the predicted helium mass fraction at the solar surface is higher, when including overshooting. {\color{black}Moreover, the convective envelope is too shallow, as is well known
for AGSS09 \citep{Serenelli2009}. This explains the deeper overshooting required in these models.}

For both compositions, all models with and without overshooting predict an decrease in the surface abundance of $^9\mathrm{Be}$ by a factor of $1.1-1.3$. While lithium is effectively destroyed at temperatures above $2.5\times10^6\,\mathrm{K}$, the destruction of beryllium sets in at temperatures exceeding $3.5\times10^6\,\mathrm{K}$ and hence requires deeper mixing. Measurements of the solar beryllium depletion are notoriously difficult, but studies seem to suggest a very small difference between the photospheric and meteoritic values \citep{Grevesse1998,Asplund2009}. Data are even consistent with beryllium being undepleted \cite{Balachandran1998}. We hence note that none of the overshooting parametrizations strongly contradict measurements of beryllium destruction. 

Moreover, according to \cite{Gloeckler1996} and \cite{Geiss1998}, the $^3\mathrm{He}/^4\mathrm{He}$ ratio has not changed by more than approximately $10\,\%$ over the last $3\,\mathrm{Gyr}$. Consequently, just as beryllium, $^3\mathrm{He}/^4\mathrm{He}$ restricts the mixing below the base of the convection zone \cite[cf.][]{Vauclair2000}. For all models, we find $^3\mathrm{He}/^4\mathrm{He}$ to increase by $3-4\,\%$, throughout the entire evolution from the pre-MS up to the present solar~age. 

\section{Helioseismic Properties}

\begin{figure}
\centering
\includegraphics[width=\linewidth]{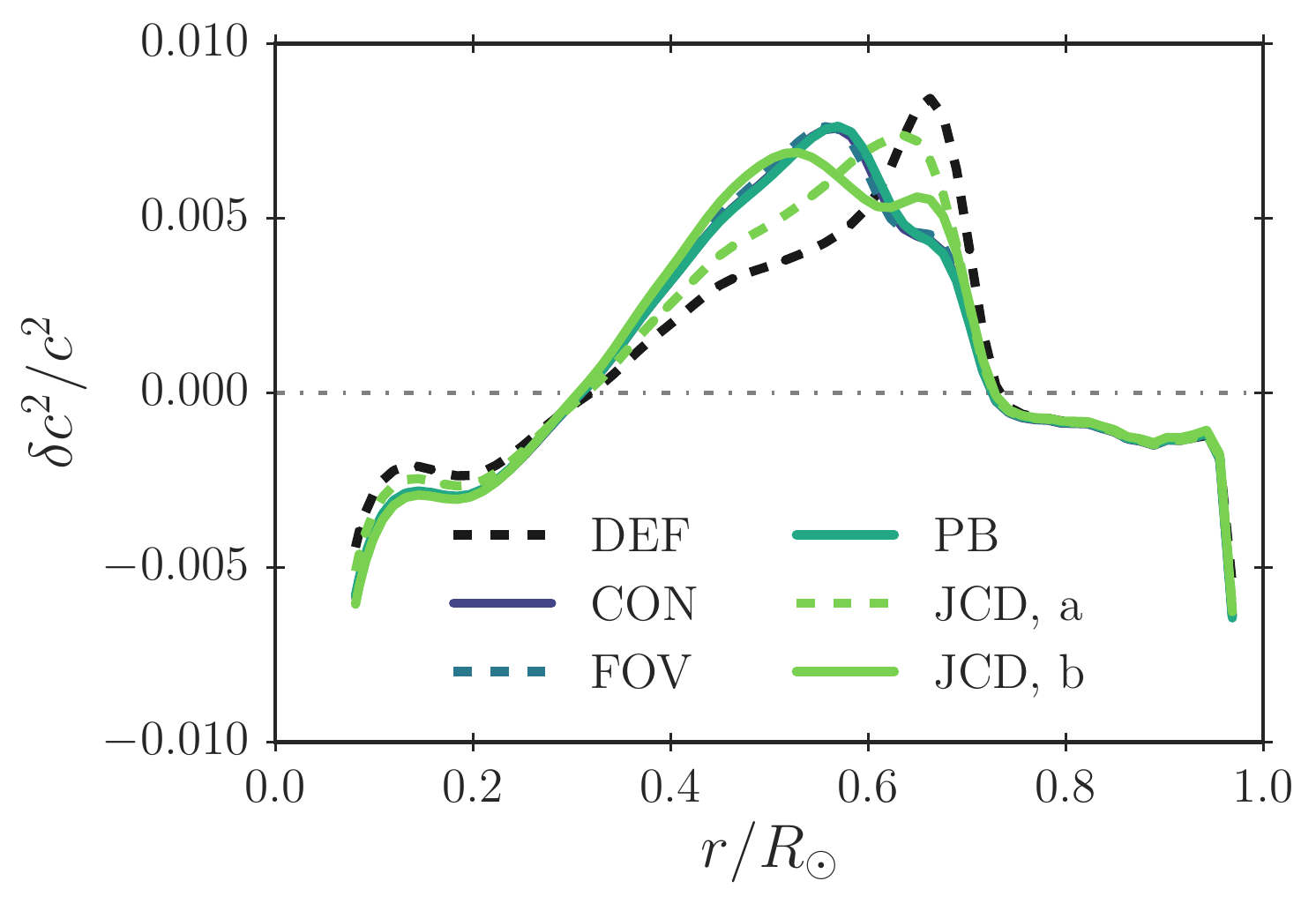}
\caption{Squared sound speed difference between models with different overshooting approaches and the Sun inferred by SOLA inversion. We use GS98. The results are based on the observed 'Best-set' frequencies by Basu et al. (1997). The associated confidence intervals of $\delta c^2/c^2$ and the centres of the averaging kernels are not included for clarity. All solar calibration models in this figure take microscopic diffusion of metals into account.
}
\label{fig:comp}
\end{figure}

The propagation of solar oscillations is determined by the adiabatic sound speed $c$:
\begin{equation}
c^2 = \frac{\Gamma_1 p}{\rho}.
\end{equation}
Here $\Gamma_1=(\partial \ln p/\partial \ln \rho)_\mathrm{ad}$ denotes the first adiabatic index,  $p$ is the pressure, and $\rho$ is the density. The solar sound speed profile can be inferred from observed oscillation frequencies by the means of the SOLA inversion technique \citep{Pijpers1992,jcd1995}. The results from inversion rely on the computation of adiabatic model frequencies, for which we have employed the Aarhus adiabatic oscillations package, \textsc{adipls} \citep{jcd2008}. Figure~\ref{fig:comp} shows the corresponding difference in the squared sound speed between the models~($c_\mathrm{mod}$) and the Sun~($c_\mathrm{sun}$):
\begin{equation}
\frac{\delta c^2}{c^2} = \frac{c_\mathrm{sun}^2-c_\mathrm{mod}^2}{c_\mathrm{sun}^2}.
\end{equation}
For every solar calibration model, for which we present the inferred sound speed difference in this paper, we have re-evaluated the sound speed profile of the Sun, based on the frequency differences between the associated model frequencies and observations \citep{Basu1997}, using the SOLA inversion technique. For this purpose, J.~Christensen-Dalsgaard has kindly provided the necessary tools and kernels.

{\color{black}In stead of inferring $\delta c^2/c^2$ by inversion, many authors compare to a previously inferred solar sound speed profile based on a reference model\footnote{{\color{black}When comparing to a pre-inferred sound speed profile of the Sun, one may include the uncertainties evaluated by \cite{Degl1997} \citep[see also][]{Vinyoles2017}. However, the fact that we repeat the inversion for each solar calibration model makes these uncertainties somewhat misleading as they are most certainly too conservative (cf. J.~Christensen-Dalsgaard, private communication, based on a discussion between J.~Christensen-Dalsgaard and A.~Serenelli). We have hence omitted these uncertainties in the plots.}}. To check our inversion results, we have compared the sound speed profile of each of our solar calibration models with the solar sound speed profile inferred by \cite{Basu2008}. From these comparisons, we draw the exact same conclusions as elaborated upon below.} 

As can be seen from the Fig.~\ref{fig:comp}, the model without overshooting shows a characteristic anomaly near the base of the convective envelope: This is the tachocline anomaly.

\begin{figure}
\centering
\includegraphics[width=\linewidth]{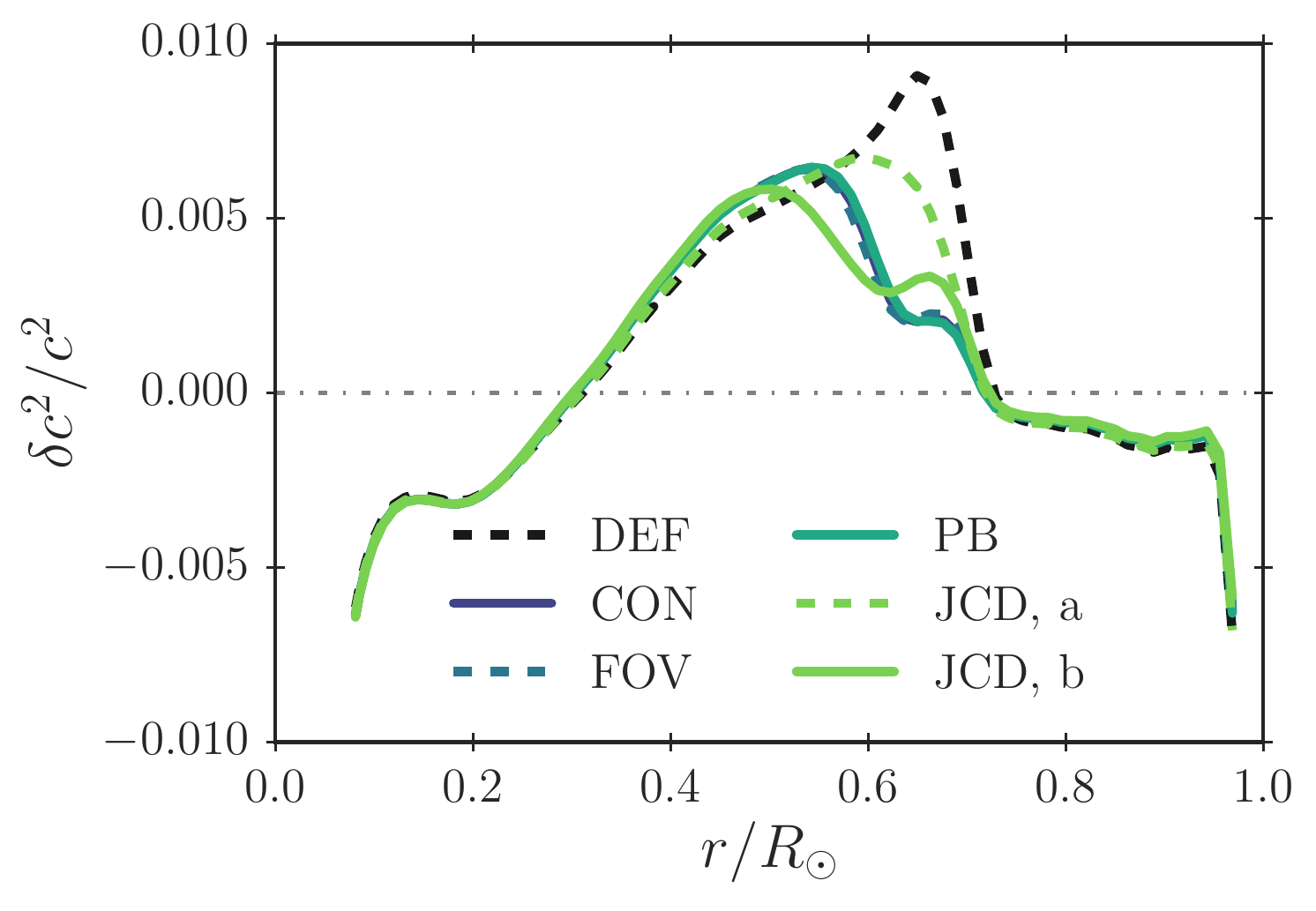}
\caption{Equivalent to Fig.~~\ref{fig:comp} but only including diffusion of H, $^4$He, $^7$Li and $^9$Be.
}
\label{fig:comp_nodiff}
\end{figure}

While overshooting {\color{black}slightly modifies} the tachocline anomaly, the implementation of overshooting alone leads to a higher discrepancy in the sound speed at lower depths (cf.~Fig.~\ref{fig:comp}), in accordance with the work published by other authors \citep[e.g.][]{jcd2011,jcd2018,Schlattl1999}. We find that this feature does not appear, if microscopic diffusion of metals (i.e. diffusion of elements other than H and $^4$He) is ignored: including the microscopic diffusion of metals improves the sound speed of the model without overshooting, throughout the radiative region, {\color{black}but} worsens the agreement, especially in the broader vicinity of the tachocline anomaly, in the case of models with overshooting. Figure~\ref{fig:comp_nodiff} shows the squared sound speed difference between the helioseismic Sun and solar calibration models, for which microscopic diffusion {\color{black}of metals other than lithium and beryllium} has been neglected. {\color{black}We note that the inclusion of the trace elements, lithium and beryllium, has no influence on the sound speed profile.} We have used the same values for the overshooting parameters as in the case of the models presented in Fig.~\ref{fig:comp}. The neglect of microscopic diffusion {\color{black}of metals} reduces the lithium depletion slightly. {\color{black}Comparing Figs~\ref{fig:comp} and \ref{fig:comp_nodiff}, it is clear that the inclusion of overshooting creates a sensitivity on metal diffusion.} 
The fact that the models in Fig.~\ref{fig:comp_nodiff} get rid of the Tachocline anomaly without introducing new anomalies in the sound speed profile is a tantalizing result{\color{black}. H}ere, we refrain ourselves from performing a detailed investigation of the treatment of diffusion, as this is beyond the scope and not the focus of the present paper.

Diffusive overshooting has been shown to lead to better agreement between the predicted sound speed and observations for lower values of the overshooting parameters involved --- see, \cite{Schlattl1999} and \cite{jcd2018} for the FOV and the JCD parametrization, respectively. These models do, however, not simultaneously solve the lithium problem.

Of course dynamical processes, such as magnetic fields and rotation, may likewise alter the solar structure, partly explaining the remaining discrepancies in the obtained sound speed profile. Alternatively, corrections to the opacities may be required, as suggested by \cite{jcd2010}.

\begin{figure}
\centering
\includegraphics[width=\linewidth]{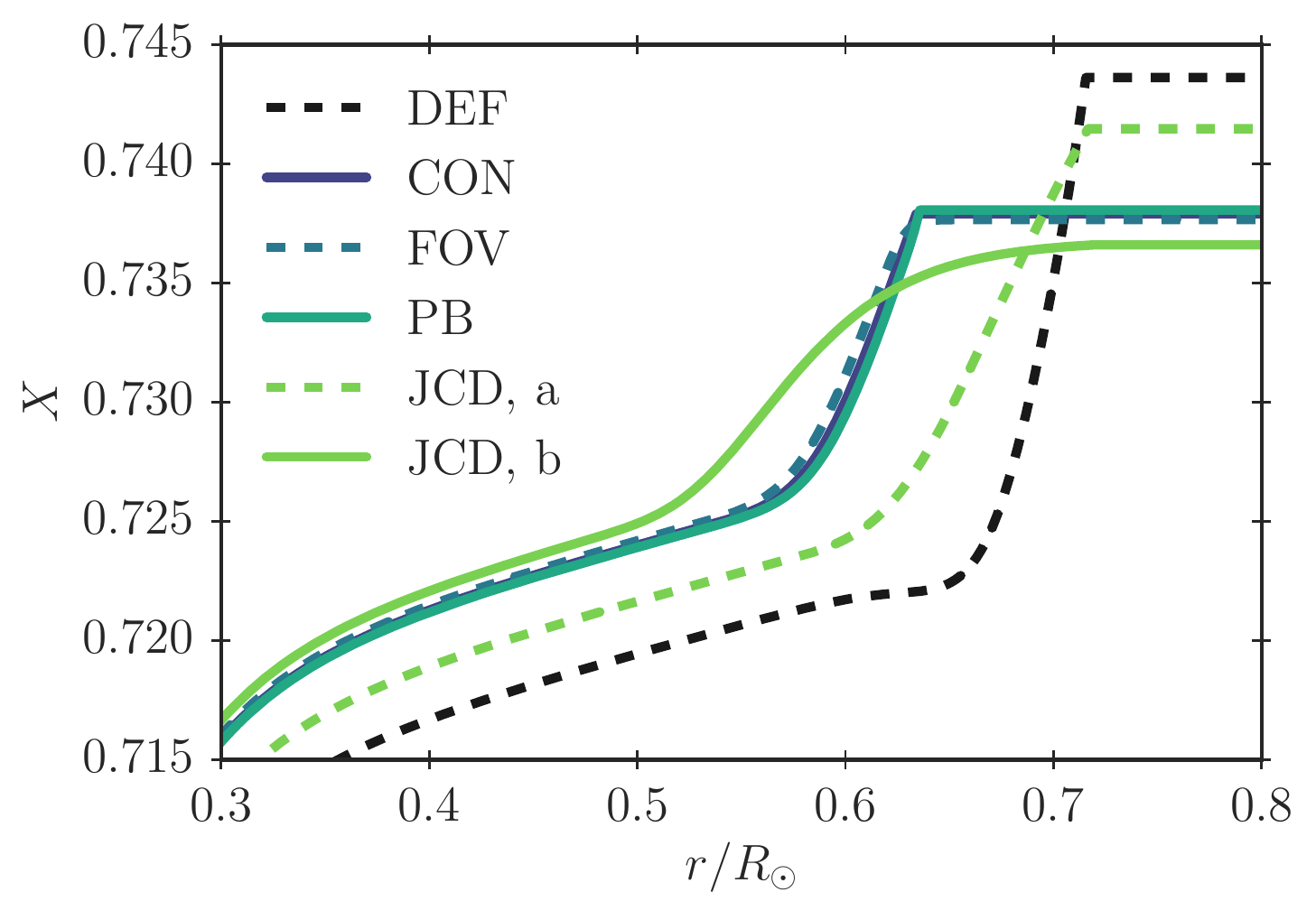}
\caption{Hydrogen abundance $X$ as a function of radius for the solar calibration models presented in Table~\ref{tab:liseis} --- i.e. the models include microscopic diffusion of metals.
}
\label{fig:X}
\end{figure}

Figure~\ref{fig:X} shows the hydrogen mass fraction $X$ as a function of radius for our solar calibration models that include microscopic diffusion of metals. The strong gradient in $X$ near the bottom of the convective zone in the model with no overshooting gives rise to the tachocline sound speed anomaly in Fig.~\ref{fig:comp}. Due to overshooting, $X$ increases, which results in an increase of the sound speed in this layer. As can be seen from Fig.~\ref{fig:X}, the JCD parametrization {\color{black}b} leads to a dramatically different hydrogen abundance profile than the other overshooting parametrizations{\color{black}, while case a lies intermediate between the default and the other cases.} However, just as the simplistic CON parametrization, the JCD parametrization is a mere toy model. Indeed, since Eq.~(\ref{eq:jcdD}) is a polynomial with three free parameters, the other suggested functional forms of $D_\mathrm{ov}(r)$ can be roughly mimicked by this equation.

{\color{black}Based on the figures discussed above we further note that all overshooting parametrizations lead to very similar chemical and thermal structures for the present Sun, if they are to reproduce the desired lithium depletion. In this respect, the parametrization does not matter.}

\begin{figure}
\centering
\includegraphics[width=\linewidth]{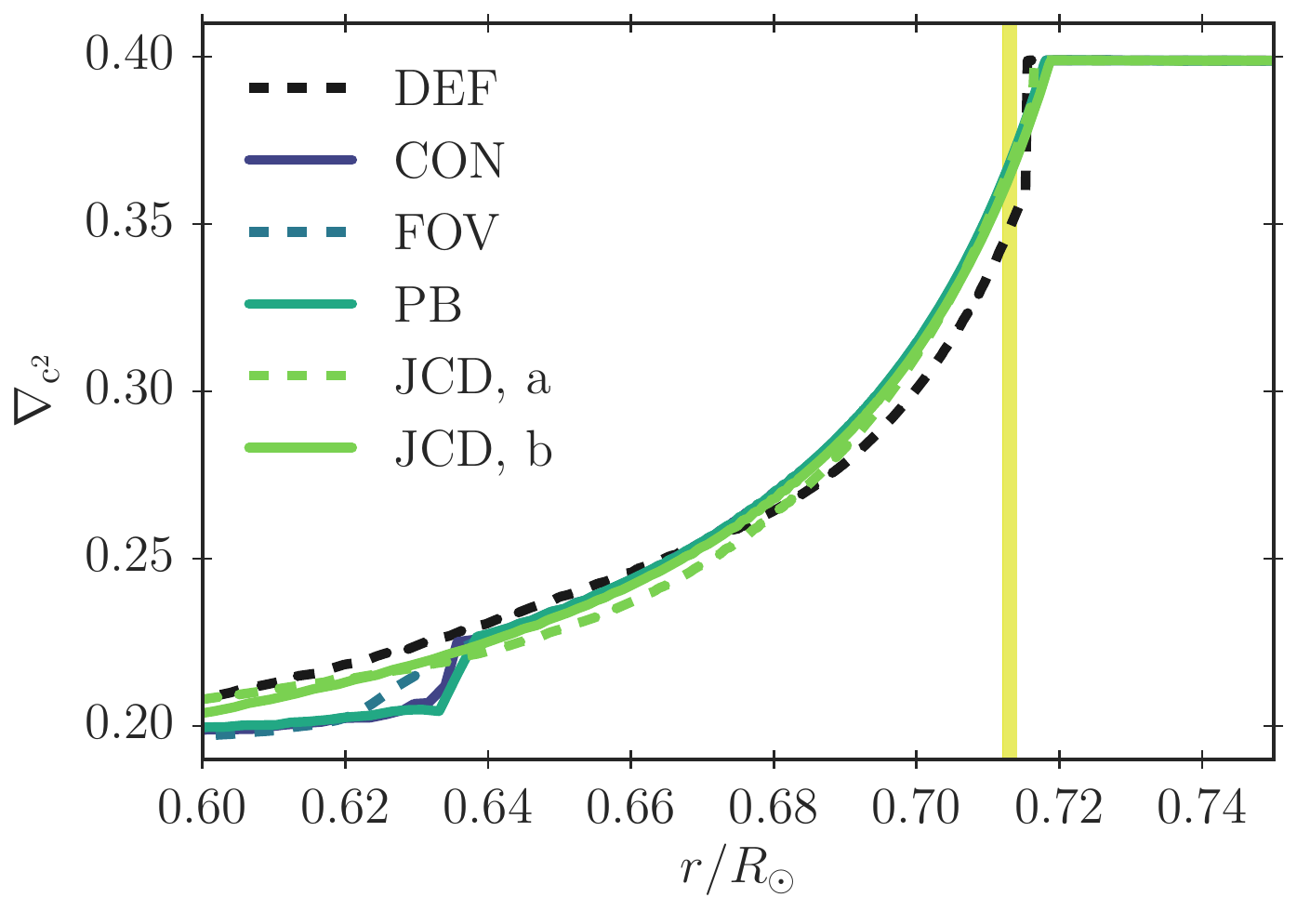}
\caption{The gradient of the squared sound speed for the solar calibration models presented in Table~\ref{tab:liseis} --- i.e. the models include microscopic diffusion of metals. The shaded area shows the seismically inferred confidence interval for the location of the base of the convective envelope found by Basu \& Antia (1997).
}
\label{fig:dc2}
\end{figure}

At the base of the solar convective envelope, there is a sharp transition in the sound speed gradient ($\nabla_{c^2}$, cf.~Fig.~\ref{fig:dc2}), due to the change in the temperature gradient associated with the transition from radiative to adiabatic heat transport \citep[cf.][]{jcd1991,jcd2011}:
\begin{equation}
\nabla_{c^2} \equiv \frac{\mathrm{d} \ln c^2}{\mathrm{d}\ln p} \approx \nabla - \frac{\mathrm{d} \ln \mu}{\mathrm{d} \ln p}, \quad \nabla \equiv \frac{\mathrm{d} \ln T}{\mathrm{d} \ln p}.
\end{equation}
Here the second equality of the first expression holds true for an ideal gas. The abrupt transition in $\nabla_{c^2}$ is referred to as an acoustic glitch. It leads to a prominent signal in the oscillation frequencies that allows for the determination of its location and hence of the base of the convection zone. This can be done, using an asymptotic absolute method, based on the Duvall relation, or employing an asymptotic differential method that relies on a reference model \citep{jcd1988,jcd1989,jcd1991}. Following the latter approach, \cite{BasuAntia1997} find the base of the convective zone to be at a radius of $0.713\pm0.001 R_\odot$. As can be seen from Fig.~\ref{fig:dc2}, the implementation of overshooting shifts the acoustic glitch outwards (cf.~Fig.~\ref{fig:dc2} {\color{black}and Tab.~\ref{tab:liseis}}), leading to a slightly worse agreement between the models and \cite{BasuAntia1997}. However, the shift is only of the order of $10^{-3}\,R_\odot$, and the acoustic glitch lies rather close to seismically predicted base of the convective envelope for all models presented in this paper with GS98. 

Furthermore, it is worth noting from Fig.~\ref{fig:dc2} that the transition in $\nabla_{c^2}$ is somewhat smoother for all models with overshooting than for the model with no overshooting --- setting $D_\textsc{jcd} = 150\, \mathrm{cm^2 \, s^{-1}}$, the transition is still rather abrupt. While a lower value of $D_\textsc{jcd}$ partly removes the tachocline anomaly \citep{jcd2018}, it does not necessarily improve all seismic properties of the model. However, in order to obtain quantitative helioseismic restrictions on the model, a detailed analysis of the oscillatory signal arising from the acoustic glitch is needed \citep[][and references herein]{jcd2011}, which is beyond the scope of this paper.

\begin{figure}
\centering
\includegraphics[width=\linewidth]{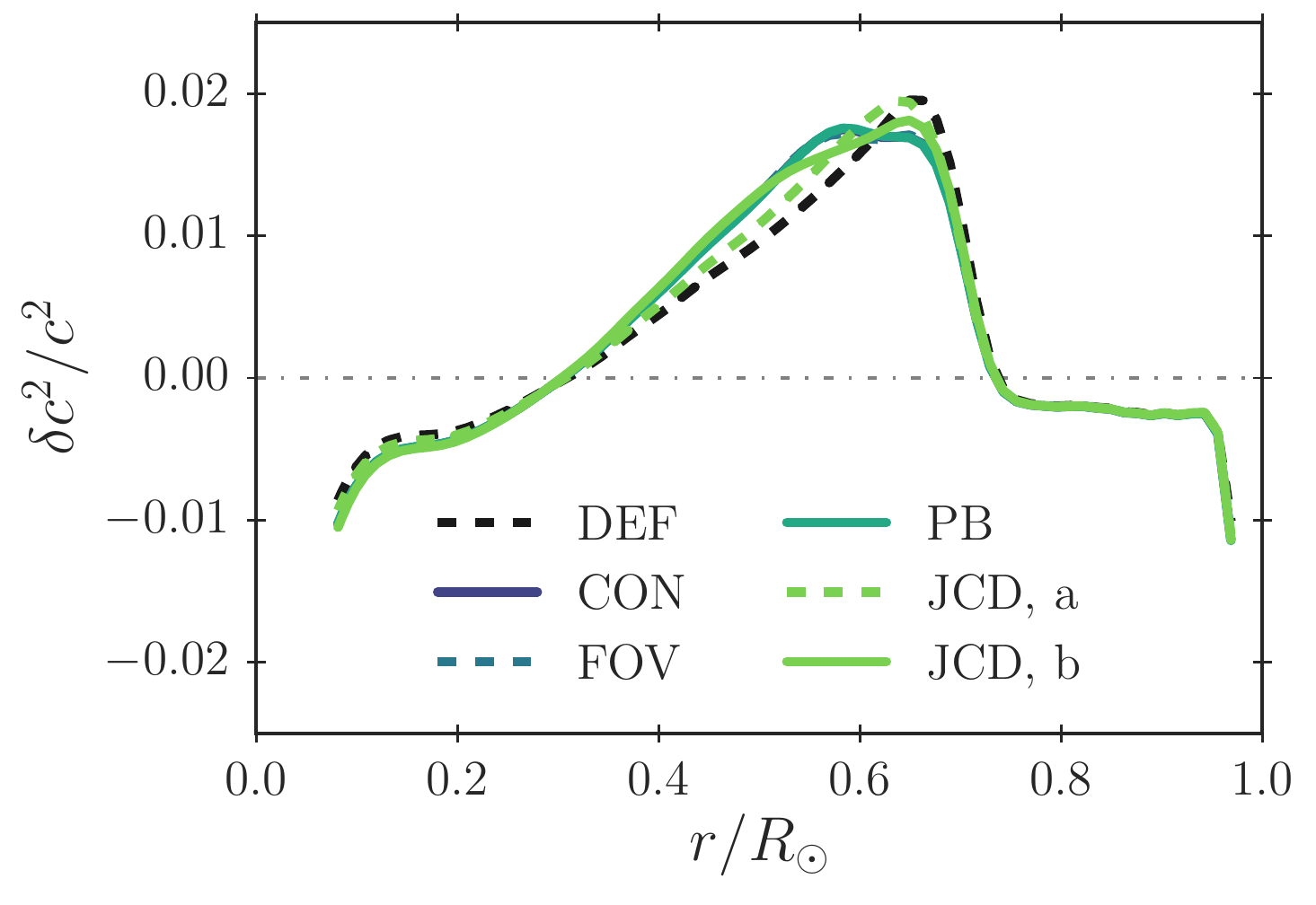}
\caption{Equivalent to Fig.~\ref{fig:comp} but for AGSS09.
}
\label{fig:comp2}
\end{figure}

For comparison, we have included the squared sound speed difference ($\delta c^2/c^2$) for solar calibration models obtained for AGSS09 in Fig~\ref{fig:comp2} (cf.~Table~\ref{tab:liseis2}). The same conclusions are drawn: Overshooting alone does not mend the inadequacies of the sound speed profile of the presented models. As in the case of GS98, the implementation of diffusive overshooting shifts the convective boundary outwards {\color{black}(cf. Tab.~\ref{tab:liseis2})} and leads to a somewhat smoother transitions in $\nabla_{c^2}$. Overall, the use of AGSS09 leads to a stronger disagreement between the model and the helioseismic Sun than the use of GS98 does. This well-known deterioration has haunted helioseismology for a decade \citep{Serenelli2009}.

\section{Conclusion}

Diffusive overshooting has been shown to either improve the predicted sound speed profile \citep[][]{jcd2007,jcd2018} or to allow for the solution of the lithium problem \citep{Baraffe2017}, when choosing suitable overshooting parameters. Discouragingly, none of the models presented in this paper solves both problems in tandem. 

For all investigated combinations of input physics, overshooting itself is not enough to obtain a perfect match between the model predictions and the inferred solar structure. In order to ensure a sufficiently high lithium depletion, each diffusive overshooting scheme introduces a new anomaly in the sound speed profile, when microscopic diffusion is taken into account. Further improvements of the input physics are needed.

Hence, the ability of a particular parametrization to solve either of these problems is itself insufficient to establish, whether said parametrization leads to a physically correct description of the overshooting layer. On the other hand, combining helioseismic measurements with information of solar abundances allows for the exclusion of some scenarios and parameter choices, imposing restrictions on additional mixing processes in the radiative zone.

This being said, all four diffusive overshooting parametrizations lead to very similar sound speed profiles, when predicting comparable evolutions of the surface lithium abundance. To this extend, the parametrization is irrelevant, i.e., on their own, the suggested diagnostic tools do not lead to a clear preference. This is not to say that all models that yield the correct surface lithium abundance are indistinguishable: As shown by \cite{Schlattl1999}, models that deplete their lithium on the pre-MS show different sound speed profiles than the models presented in this paper.

In order for the CON, FOV and PB parametrizations given by Eq.~(\ref{eq:constD})-(\ref{eq:pratt}) to recover the observed trend of lithium depletion, a suppression of overshooting is required during the early stellar evolutionary stages. In accordance with \cite{Baraffe2017}, we hence assume overshooting to be far less effective during the initial evolution, when using the PB parametrization. We take this into account by limiting the width of the overshooting layers as suggested by \cite{Baraffe2017}.
As argued by several authors, this may be justified based on rotation \citep[e.g.][]{Ziegler2003,Brummell2007,Brun2017}. Since the underlying physical parameters describing rotation are insufficiently constrained, however, we have restricted ourselves to a more simplistic approach, when dealing with the CON and FOV parametrizations, in order to reduce the number of free parameters: we completely neglect overshooting for the first $10^8\,$yr of evolution. Additional constraints on stellar rotation would render this simplification obsolete and may potentially favour one of the diffusive overshooting schemes presented here, since there is also evidence for additional mixing in young clusters \citep{Baraffe2017,Bouvier2017}. However, this is beyond the scope of the present paper.

We find that the overshooting parameters involved are rather sensitive to the input physics. It is therefore not clear, to which extend the obtained parameter values from a solar calibration can be applied in the analyses of other stars without the necessity of tuning, even if the lithium problem and the tachocline anomaly were to be solved simultaneously. Likewise, it is unclear, whether the obtained parameter values lead to an apposite description of lithium depletion at later evolutionary stages of the Sun. Simulations and observations that yield solid restrictions of and physical justifications for the employed overshooting parameters are hence much needed.

\section*{Acknowledgements}

We thank J.~Christensen-Dalsgaard, I.~Baraffe, and our referee for their collaboration, input and insights. {\color{black}We thank J.~Christensen-Dalsgaard for kindly providing the inversion tools that were used to infer the solar sound speed profile for each of our solar calibration models. He provided indispensable guidance as well as all necessary kernels and settings.}
Furthermore, we record our gratitude to IMPRS on Astrophysics at the Ludwig-Maximilians University as well as the funding we received from the Max-Planck Society. 












\bsp	
\label{lastpage}
\end{document}